\documentclass[twoside,journal]{IEEEtran}
\usepackage{amsfonts}
\usepackage{amssymb}
\usepackage{amsmath}
\usepackage{mathrsfs}
\usepackage{verbatim}
\usepackage{subfigure}
\usepackage{balance}
\usepackage{booktabs}
\usepackage{fancybox}
\usepackage{bm}
\usepackage{changepage}
\usepackage{extarrows}
\usepackage{algorithm}
\usepackage{algorithmic}
\usepackage{multirow}
\usepackage{array}
\usepackage{graphicx}
\usepackage{epstopdf}
\usepackage{caption}
\usepackage{color}
\begin{document}

\title{\LARGE Improving Anti-Eavesdropping Ability without Eavesdropper's CSI: A Practical Secure Transmission Design Perspective}
\author{
	Tong-Xing Zheng,~\IEEEmembership{Member,~IEEE},~
    Hui-Ming Wang,~\IEEEmembership{Senior~Member,~IEEE},~
Hao Deng
\thanks{
T.-X. Zheng and H.-M. Wang are with the School of Electronic and Information Engineering, Xi'an Jiaotong University, and also with the Ministry of Education Key Laboratory for Intelligent Networks and Network Security, Xi'an, Shaanxi 710049, P. R. China (e-mail: zhengtx@mail.xjtu.edu.cn; xjbswhm@gmail.com).
}
\thanks{H. Deng is with the School of Physics and Electronics, Henan University, Kaifeng, Henan 475000, P. R. China (e-mail: gavind@163.com).}
}

\maketitle
\vspace{-0.8 cm}

\begin{abstract}
This letter studies the practical design of secure transmissions without knowing eavesdropper's channel state information (ECSI). An ECSI-irrelevant metric is introduced to quantize the intrinsic \emph{anti-eavesdropping ability} (AEA) that the transmitter has on confronting the eavesdropper via secrecy encoding together with artificial-noise-aided signaling. Non-adaptive and adaptive transmission schemes are proposed to maximize the AEA with the optimal encoding rates and power allocation presented in closed-form expressions. Analyses and numerical results show that maximizing the AEA is equivalent to minimizing the secrecy outage probability (SOP) for the worst case by ignoring eavesdropper's receiver noise. 
Therefore, the AEA is a useful alternative to the SOP for assessing and designing secure transmissions when the ECSI cannot be prior known.
\end{abstract}

\vspace{-0.4 cm}
\IEEEpeerreviewmaketitle

\section{Introduction}

\IEEEPARstart{P}{hysical} layer security (PLS), emerging as a promising paradigm for achieving perfect wireless secrecy against eavesdropping attacks, has attracted a stream of research during the past decade \cite{Wang2016Physical_book}. 
In general, researchers use metrics such as secrecy capacity \cite{Wyner1975Wire-tap}, ergodic secrecy capacity \cite{Goel2008Guaranteeing} and secrecy outage probability (SOP) \cite{Zhang2013Design}-\cite{Zheng2017Safeguarding}, to assess the secrecy level of physical-layer transmissions for different scenarios. 
{As can be seen, these metrics are highly relevant to eavesdropper's (instantaneous or statistic) channel state information (ECSI). 
Actually, in a real wiretap scenario, since the eavesdropper is usually passive, even that detecting it is difficult, let alone acquiring its CSI. 
In this regard, it is unrealistic to use ECSI-relevant metrics to guide the design of secure transmissions.}

In order to bridge the gap between the theoretical research of PLS and its practical application, this letter studies the design of secure transmissions without knowing the ECSI.
When the ECSI is lacking, a reasonable way to enhance the secrecy would be to exploit all the available resources to make eavesdropping difficult while guaranteeing a certain level of quality of service (QoS), e.g., high transmission reliability and low delay, etc. 
Inspired by this, we introduce a new ECSI-irrelevant metric to quantize the \emph{anti-eavesdropping ability} (AEA) of a secure transmission system, namely, the intrinsic ability that the transmitter can exert negative impacts on the eavesdropper via secrecy encoding in conjunction with artificial-noise-aided (AN-aided) signaling.
Our primary goal is to reach a maximum AEA subject to certain QoS constraints. To this end, we design both non-adaptive and adaptive transmission schemes based on the instantaneous and statistic CSIs of the legitimate channel, respectively. 
In each scheme, we maximize the AEA by jointly optimizing the encoding rates and the power allocation between the information signal and the AN. 
We establish an analytical relationship between the AEA and SOP, and reveal the equivalence between maximizing the AEA and minimizing the SOP for a worst-case scenario where eavesdropper's receiver noise is ignored.

We point out that several studies have also discussed the design of secure transmissions without ECSI  \cite{Swindlehurst2009Fixed}-\cite{He2017Joint} or with estimated ECSI \cite{Zhou2015Secure,Mei2017Energy}. They have mainly focused on the design of secrecy signaling, but encoding rates are not part of their concern. 
However, in our work we consider a combined design from the perspectives of secrecy channel encoding and signaling with the aid of an ECSI-irrelevant metric.

\section{System Model and Problem Description}
We consider a secure transmission from an $N$-antenna transmitter Alice to a single-antenna receiver Bob over a Rayleigh fading channel in the presence of an eavesdropper Eve. 
{The channel coefficient vectors of Bob and Eve are respectively denoted by ${\mathbf h}_b, {\mathbf h}_e\in \mathbb{C}^{N\times1}$.
We assume that the CSI of ${\mathbf h}_b$ is known at Alice, but that of ${\mathbf h}_e$ is not.
We also assume that the entries ${ h}_{b,i}$ of ${\mathbf h}_b$ and the receiver noise $n_b$ at Bob are independent and identically distributed (i.i.d.), obeying the circularly symmetric complex Gaussian distribution with zero mean and unit variance, i.e., ${ h}_{b,i}, n_b\sim\mathcal{CN} \left(0,1\right)$. }

For the purpose of secrecy, we adopt Wyner's secrecy encoding scheme \cite{Wyner1975Wire-tap} along with the null-space AN-aided signaling technique \cite{Goel2008Guaranteeing} to confound Eve. For secrecy encoding, we insert redundant information into confidential information, and deliver the two parts with rates $R_e$ and $R_s$, respectively. The rate of the entire codeword is thus $R_t=R_s+R_e$. 
Denote the channel capacities of Bob and Eve as  $C_b=\log_2(1+\gamma_b)$ and $C_e=\log_2(1+\gamma_e)$, where $\gamma_b$ and $\gamma_e$ are the respective signal-to-interference-plus-noise ratios (SINRs). If $C_b$ is less than $R_t$, a communication outage happens and Bob cannot recover secret information. If $C_e$ exceeds $R_e$, secret information might be decoded by Eve and a secrecy outage event is said to have occurred. The SOP is defined as the probability that the SINR $\gamma_e$ lies above a threshold $\beta_e \triangleq 2^{R_e}-1$, i.e.,  $p_{so}\triangleq\mathbb{P}\{\gamma_e>\beta_e\}$ \cite{Zhang2013Design}.  
As to secrecy signaling, we inject isotropic AN into the null space of the legitimate channel such that the transmitted signal is designed in the form of
\begin{equation}\label{signal_model}
{\bf x} = \sqrt{ \phi P}{\bf w} s + \sqrt{(1-\phi)P/(N-1)}{\bf Gv}.
\vspace{-0.0 cm}\end{equation}
In \eqref{signal_model}, $s$ is the information-bearing signal with $\mathbb{E}[|s|^2]=1$ and ${\bf v}\sim\mathbb{C}^{(N-1)\times 1}$ is an AN vector with i.i.d. entries $v_i\sim\mathcal{CN}\left(0,{1}\right)$; $P$ is the total transmit power of Alice subject to a power budget, i.e., $P\le P_{max}$; $\phi\in[0,1]$ is the power allocation ratio such that the power of the transmitted signal and the AN can be given by $P_S = \phi P$ and  $P_A=(1-\phi)P$, respectively;
${\mathbf{w}}={\mathbf{h}_b^{\dag}}/{\|\mathbf{h}_b\|}$ is the beamforming vector for Bob, and ${\bf{G}}\in{\mathbb{C}}^{N\times (N-1)}$ is an AN weight matrix projecting onto the null space of 
$\mathbf{h}_b$ such that
$\mathbf{h}_b^{\mathrm{T}} \mathbf{G} = \mathbf{0}$, meaning that the AN does not interfere with Bob.
{The SINRs of Bob and Eve can be respectively given by 
\begin{equation}\label{sinr_b}
\gamma_b=P_S \|{\mathbf h}_b\|^2,
\end{equation}
\begin{equation}\label{sinr_e}
\gamma_e = \frac{ P_S|\mathbf{h}_e^{\mathrm{T}} \mathbf{w}|^2}{P_A\|\mathbf{h}_e^{\mathrm{T}} \mathbf{G}\|^2/(N-1)+\sigma_e^2},
\end{equation}
where $\sigma_e^2$ denotes the receiver noise power at the Eve side. 

Note that Eve's noise power $\sigma_e^2$ is typically unknown to Alice, then a robust approach is to design for the worst-case scenario by ignoring Eve's noise, i.e., $\sigma_e^2=0$ \cite{Zhang2013Design}.
Hence, the SOP can be calculated from \eqref{sinr_e}, given below, 
\begin{equation}\label{sop}
p_{so}=\mathbb{P}\left\{\gamma_e>\beta_e\right\}=\mathbb{P}\left\{\frac{ P_S|\mathbf{h}_e^{\mathrm{T}} \mathbf{w}|^2}{P_A\|\mathbf{h}_e^{\mathrm{T}} \mathbf{G}\|^2}>\frac{\beta_e}{N-1}\right\}.
\end{equation} 
Since the wiretap channel $\mathbf{h}_e$ can be arbitrarily distributed, we cannot assess the exact SOP. Nevertheless, we can rewrite \eqref{sop} as $p_{so}=\mathbb{P}\left\{(N-1){ |\mathbf{h}_e^{\mathrm{T}} \mathbf{w}|^2}/{\|\mathbf{h}_e^{\mathrm{T}} \mathbf{G}\|^2}>{P_A\beta_e/P_S}{}\right\}$. Interestingly, regardless of the distribution of $\mathbf{h}_e$, minimizing the SOP is equivalent to maximizing the following metric in the worst-case scenario,\footnote{As we will show in the simulations, the maximum AEA actually yields an SOP approaching the real minimum SOP for a more general case. }
\begin{equation}\label{aea_def}
\Omega\triangleq \frac{P_A}{P_S}\beta_e.
\end{equation}

This metric formally describes the collective negative impact imposed on Eve: $\beta_e$ captures the confusion created against illegal decoding via secrecy encoding; ${P_A}/{P_S}$ captures the degradation of the wiretap channel via AN-aided signaling. 
The metric $\Omega$ itself has a specific physical significance, i.e., it quantizes Alice's intrinsic ability in terms of anti-eavesdropping. In this sense, we refer to the metric $\Omega$ as \emph{AEA}.
Distinguished from the SOP that relies heavily on the knowledge of Eve's statistic CSI, the AEA is completely irrelevant to any parameters regarding to Eve. In this regard, the AEA helps to guide an easy-to-implement PLS approach for practical applications as it bypasses the assumption of the availability of ECSI. In addition, once the distribution of $\mathbf{h}_e$ is prior known, the AEA can also be used to evaluate the SOP. 

We emphasize that with the metric AEA, the SOP can be interpreted as the probability that the power gain of the signal over the AN at Eve exceeds the AEA. Such a close relationship actually reveals \emph{a fundamental principle of lifting the secrecy level in the absence of ECSI, that is, in order to enhance the secrecy we should maximize the AEA while guaranteeing a reliable link to the intended receiver.}
Inspired by this, in this letter we will tackle the problem of maximizing the AEA.  

Recalling the secrecy encoding where $R_e = R_t - R_s$, we have $\beta_e = (\beta_t-\beta_s)/(1+\beta_s)$ with $\beta_t=2^{R_t}-1$ and $\beta_s=2^{R_s}-1$. Hence, the AEA in \eqref{aea_def} can be rewritten as 
\begin{equation}\label{aea_def2}
\Omega= \frac{1-\phi}{\phi}\frac{\beta_t-\beta_s}{1+\beta_s}.
\end{equation}

In order to avoid an undesired communication outage (i.e., $\beta_t>\gamma_b$) or an intolerable high risk of secrecy outage (i.e., $\beta_e<\gamma_e$), we follow an on-off criterion where Alice transmits only when the channel gain $\|{\mathbf h}_b\|^2$ exceeds a pre-established threshold $\mu$; otherwise the transmission is suspended. 
Hence, Alice has a transmission probability 
$p_t= \mathbb{P}\left\{\|{\mathbf h}_b\|^2>\mu\right\}$, which is directly related to the average delay, i.e., the larger $p_t$ the shorter the expected delay \cite{Zhang2013Design}.
{Besides, due to the Rayleigh fading, $\|{\mathbf h}_b\|^2$ is a normalized gamma variable with the shape parameter $N$,} and then $p_t$ can be computed as
\begin{equation}\label{pt}
p_t =\mathbb{P}\left\{\|{\mathbf h}_b\|^2>\mu\right\}=e^{-\mu}\sum_{k=0}^{N-1}\frac{\mu^k}{k!}.
\end{equation}

Since the encoding rates and the power allocation can be dynamically adjusted according to channel realization, parameters $\phi$, $\beta_t$ and $\beta_s$ are potentially functions of ${\mathbf h}_b$.
Therefore, the overall AEA under the on-off strategy can be given by 
\begin{equation}\label{overall_aea}
\bar{\Omega}=\mathbb{E}_{{\mathbf h}_b}\left[\Omega(\phi, \beta_t, \beta_s)~|~\|{\mathbf h}_b\|^2>\mu\right].
\end{equation}

In the following sections, we aim at maximizing the overall AEA $\bar{\Omega}$ while guaranteeing a minimum transmission probability $\delta$, i.e., $p_t\ge\delta$, and a minimum secrecy rate $R_m$, i.e., $R_s\ge R_m$, or equivalently, $\beta_s\ge\beta_m\triangleq 2^{R_m}-1$.

\section{Non-Adaptive Secure Transmission Scheme}
This section studies a non-adaptive secure transmission (NAST) scheme where all the parameters are designed based on the statistic CSI of ${\mathbf h}_b$ and remain unchanged during the transmission period. This scheme can be accomplished in an off-line manner with a low  implementation complexity. 

It is obvious that in the NAST scheme the AEA, denoted as  $\Omega_{nast}$, is independent of $\mathbf{h}_b$. Hence, the conditioned expectation in \eqref{overall_aea} can be ignored and the overall AEA $\bar{\Omega}_{nast}$ is described as \eqref{aea_def2}. {Since Alice transmits if $\|\mathbf{h}_b\|^2$ exceeds $ \mu$, the condition $\beta_t\leq\phi P\mu<\phi P \|\mathbf{h}_b\|^2 $ should be satisfied for achieving a reliable transmission. Therefore, the problem of maximizing the AEA can be formulated as follows:
\begin{subequations}
	\begin{align}\label{aea_max_na}
	\max_{P,\phi, \beta_t,\beta_s,\mu}~  &{\Omega_{nast}=\frac{1-\phi}{\phi}}\frac{\beta_t-\beta_s}{1+\beta_s}\\
	\label{aea_max_n1}
	~~\mathrm{s.t.}~~& \beta_m<\beta_s\leq \beta_t\leq\phi P\mu,\\
	\label{aea_max_n2}
	~~&p_t \ge\delta,\\
	\label{aea_max_n3}
	~~&0< P\le  P_{max},\\
	\label{aea_max_n4}
	~~&0<\phi<1.
\end{align}
\end{subequations}

In what follows, we will solve the multi-variable problem given above by successively designing the optimal parameters, considering both multi- and single-antenna transmitter cases.

\subsection{Multi-antenna Transmitter Case}
The objective function in \eqref{aea_max_na} indicates that in order to maximize $\Omega_{nast}$, we should first set $\beta_t$ to its maximum, which is $\phi P\mu$ from \eqref{aea_max_n1}. Substituting $\beta_t^*=\phi P\mu$ into \eqref{aea_max_na}, we find that for a given $\phi$, $\Omega_{nast}$ decreases with $\beta_s$ and increases with $P$ and $\mu$, meaning that $\Omega_{nast}$ is maximized at $\beta_s^*=\beta_m$, $P^*=P_{max}$ and $\mu^* = \mu_N^o$, where $\mu_N^o$ denotes the maximum of $\mu$. 
Since $p_t$ decreases with $\mu$, $\mu_N^o$ should make the constraint \eqref{aea_max_n2} active, i.e., $\mathbb{P}\left\{\|{\mathbf h}_b\|^2>\mu_N^o\right\}=\delta$, and the value of $\mu_N^o$ can be efficiently calculated using the bisection method.

Substituting the optimal $\beta_t^*$ and $\beta_s^*$ into \eqref{aea_def2} yields
\vspace{-0.0 cm}\begin{equation}\label{aea_na}
	\Omega_{nast}(\phi)= \frac{1-\phi}{\phi}\frac{\phi P_{max}\mu_N^o -\beta_m}{1+\beta_m}.
	\end{equation}
	Apparently, to achieve a positive $\Omega_{nast}(\phi)$, the minimum $\delta$ and $\beta_m$ must satisfy the following inequality
	\begin{equation}\label{pt_rm}
P_{max}\mu_N^o>\beta_m.
	\end{equation}	
	After some algebraic operations, we can prove $\Omega_{nast}(\phi)$ concave on $\phi$, and the maximum $\Omega_{nast}(\phi)$ is achieved at $d\Omega_{nast}(\phi)/d\phi=0$ with the optimal $\phi^*=\sqrt{\beta_m/(P_{max}\mu_N^o)}$. The overall maximum AEA $	\bar\Omega_{nast}^*$ can be eventually given by
	\begin{equation}\label{max_aea_na_mul}
	\bar\Omega_{nast}^*=\Omega_{nast}^* = \frac{\left(\sqrt{P_{max}\mu_N^o}-\sqrt{\beta_m}\right)^2}{1+\beta_m}.
	\end{equation} 
\subsection{Single-antenna Transmitter Case}
If Alice has a single transmit antenna, the null-space AN can no longer be realized and the leaked AN will also interfere with Bob. 
In this case, the SINR of Bob changes to $\gamma_b=\phi P|h_b|^2/\left((1-\phi) P|h_b|^2+1\right)$. 
We can formulate the AEA maximization problem similarly as problem \eqref{aea_max_na}, simply with constraints \eqref{aea_max_n1} and \eqref{aea_max_n2} modified to $\beta_t\le \phi P\mu/\left[(1-\phi) P\mu+1\right]$ and $p_t = \mathbb{P}\left\{|h_b|^2>\mu\right\}=e^{-\mu}$, respectively.
Afterwards, we can successively determine the optimal parameters, namely, $P^*=P_{max}$, $\mu^*=\mu_1^o\triangleq\ln\frac{1}{\delta}$, $\beta_s^*=\beta_m$, $\beta_t^* = \sqrt{P_{max}\mu_1^o\beta_m}$, and $\phi^*=\sqrt{\frac{\beta_m}{P_{max}\mu_1^o}}\frac{1+P_{max}\mu_1^o}{1+\sqrt{{P_{max}\mu_1^o\beta_m}}}$.  
The condition for achieving a positive AEA is $P_{max}\mu_1^o>\beta_m$, and the resulting overall AEA is given by 
	\begin{equation}\label{max_aea_na_sin}
	\bar\Omega_{nast}^* =\Omega_{nast}^* = \frac{\left(\sqrt{P_{max}\mu_1^o}-\sqrt{\beta_m}\right)^2}{(1+\beta_m)\left(1+P_{max}\mu_1^o\right)}.
	\end{equation}

The design procedure above shows a strong coupling between the encoding rates and power allocation. This proves the necessity of combining secrecy channel encoding and signaling in guaranteeing wireless secrecy.  
The obtained solutions give some useful insights into both system performance and design guidelines for practical secure transmissions.

1) For the sake of anti-eavesdropping, Alice should transmit with full power, even though this would strengthen Eve's signal. This is because only  higher transmit power can support a larger codeword rate such that a larger redundant rate can be attained against eavesdropping. Likewise, a small secrecy rate beyond the required minimum one is preferred.

2) We can prove the optimal power allocation $\phi^*$ decreasing with $N$ by realizing that $\mu_N^o$ increases with $N$.
This indicates that higher power is allowed to be devoted to the AN injection with more transmit antennas. Besides, $\phi^*$ increases with $R_s$ since higher power should be ensured for sending the useful signal to support the target secrecy rate. 

3) There exists a non-trivial trade-off between secrecy and delay under the on-off strategy. The condition of a positive AEA shows that the requirements of a low transmission delay (i.e., a large $\delta$) and a large secrecy rate $R_m$ cannot be satisfied simultaneously. In other words, enhancing the secrecy performance is at the cost of increasing the transmission delay. 

4) Comparing the AEAs in \eqref{max_aea_na_mul} and \eqref{max_aea_na_sin}, we observe that apart from an array gain loss (as $\mu_N^o>\mu_1^o$), the single-antenna case suffers a performance loss arisen from the leaked AN, i.e., ${1}/{\left(1+P_{max}\mu_1^o\right)}$. This highlights the superiority of multi-antenna techniques in improving wireless secrecy. In addition, as power budget $P_{max}$ increases, the AEA in \eqref{max_aea_na_mul} increases continuously whereas that in \eqref{max_aea_na_sin} tends to a constant $1/(1+\beta_m)$ which is bottlenecked by the required secrecy rate $R_m$.

\section{Adaptive Secure Transmission Scheme}
To improve the AEA further, this section studies an adaptive secure transmission (AST) scheme where the fading status of channel $\mathbf{h}_b$ is adequately exploited. In this case, all the parameters are adaptively adjusted according to the instantaneous CSI of $\bf{h}_b$, that is, they are functions of $\mathbf{h}_b$.

Recalling \eqref{overall_aea}, we find that for a given threshold $\mu$, if we maximize $\Omega_{ast}(\mathbf{h}_b)$ under an arbitrary $\mathbf{h}_b$, the overall AEA $\bar{\Omega}_{ast}$ is naturally maximized. Therefore, we can first focus on maximization of $\Omega_{ast}(\mathbf{h}_b)$ for a fixed $\mu$, and then determine the optimal $\mu$ to maximize $\bar{\Omega}_{ast}$.
Since Alice knows ${\mathbf h}_b$, she can set codeword rate $R_t$ to the channel capacity of the legitimate channel $C_b$, i.e., $\beta_t=\gamma_b=\phi P \|{\mathbf h}_b\|^2$. 
The AEA maximization problem is thus formulated as follows:
\begin{subequations}
	\begin{align}\label{aea_max_a}
	\max_{P,\phi,\beta_s}~  &{\Omega_{ast}({{\mathbf h}_b})=\frac{1-\phi}{\phi}}\frac{\gamma_b-\beta_s}{1+\beta_s}\\
	\label{aea_max_c1}
	~~\mathrm{s.t.}~~& \beta_m\le\beta_s<\gamma_b,\\
	\label{aea_max_c2}
	~~&0<\phi<1,\\
	\label{aea_max_c3}
	~~&0<P\le P_{max}.
	\end{align}
\end{subequations}

Since problems \eqref{aea_max_a}  and \eqref{aea_max_na} share a unified optimization procedure for a fixed $\mu$, the detailed design is omitted here due to space limitation. We finally obtain the optimal parameters in the AST scheme by replacing $\mu_N^o$ in the NAST case with $\|{\mathbf h}_b\|^2$. 
Understanding that $\Omega_{ast}^*({\mathbf h}_b)$ increases with $\|{\mathbf h}_b\|^2$, we can prove from \eqref{overall_aea} that the overall AEA is maximized if threshold $\mu$ reaches the maximum value, i.e., $\mu^* = \mu_N^o$ with $\mu_N^o$ defined previously. Similar conclusions can be developed as in the NAST case. For example, the power resource should be completely utilized to maximize the AEA. In addition to the delay-secrecy trade-off, a rate-secrecy trade-off is captured in power allocation, e.g., increasing AN power benefits the secrecy but might fail to support a high secrecy rate. Only when the quality of the legitimate channel is good enough (a large $\|{\mathbf h}_b\|^2$) and meanwhile the secrecy rate constraint is moderate (a small $R_m$), can higher AN power be permitted for degrading the wiretap channel.

Substituting the optimal $\Omega_{ast}^*({\mathbf h}_b)$ and $\mu^*$ into \eqref{overall_aea} yields the maximum overall AEA. 
As expected, the positive AEA condition $\|{\mathbf h}_b\|^2>\mu_N^o>\beta_m/P_{max}$ always yields $\Omega_{ast}^*({\mathbf h}_b)>\Omega_{nast}^*$. This indicates, due to an adaptive design, the AST scheme outperforms the NAST scheme in terms of the overall AEA, i.e., $\bar\Omega_{ast}^*>\bar\Omega_{nast}^*$. 
However, in the single-antenna case when $P_{max}$ or $|h_b|^2$ increases to infinity,  $\Omega_{ast}^*({ h}_b)$ approaches the same constant as $\Omega_{nast}^*$, i.e., $1/(1+\beta_m)$ .

\section{Numerical Results}
This section presents numerical results on the AEAs as well as the corresponding SOPs. The comparison between the SOPs with and without the ECSI is also provided.

For simplicity, we consider a single-antenna Eve with Rayleigh fading channels. After some algebraic operations, we express the SOP $p_{so}({\mathbf h}_b)$ as a function of the AEA $\Omega({\mathbf h}_b)$, 
\begin{align}\label{pso}
&p_{so}({\mathbf h}_b)=\begin{cases}
~e^{-\frac{\sigma_e^2\Omega({\mathbf h}_b) }{P(1-\phi)(1-\Omega({\mathbf h}_b))}},&N= 1,\\
~ e^{-\frac{\sigma_e^2\Omega({\mathbf h}_b)}{P(1-\phi)}}\left(1+\frac{\Omega({\mathbf h}_b)}{N-1}\right)^{1-N}, &N \ge 2.
\end{cases}
\end{align}
Using the optimal parameters obtained previously, the minimum overall SOP can be given by 
\begin{equation}
\bar{p}_{so}^*=\mathbb{E}_{{\mathbf h}_b}\left[p^*_{so}({\mathbf h}_b)~|~\|{\mathbf h}_b\|^2>\mu^*\right].
\end{equation}

\begin{figure}[!t]
	\centering
	\includegraphics[width = 3.0in]{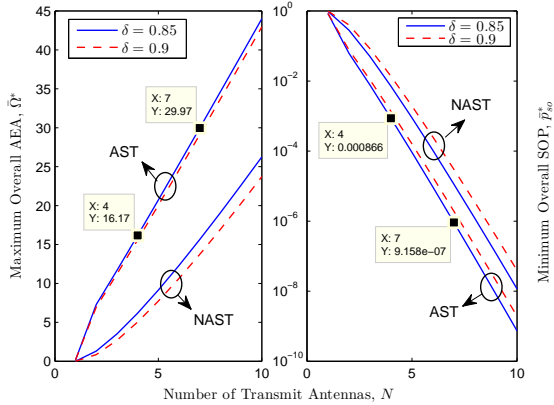}
	\caption{Maximum overall AEA $\bar \Omega^*$ and minimum overall SOP $\bar{p}_{so}^*$ vs. $N$ and $\delta$, with $P_{max}=10$ dBm and $\beta_m=1$.}
	\label{AEA_SOP_N_DELTA}
\end{figure}
Fig. \ref{AEA_SOP_N_DELTA} plots the maximum overall AEA $\bar \Omega^*$ and the corresponding minimum overall SOP $\bar p^*_{so}$ as functions of $N$ for different values of $\delta$. 
Simulation results match well with the theoretical values, verifying the correctness of our analyses.
A consistent one-to-one match between $\bar \Omega^*$ and $\bar p^*_{so}$ is shown. The delay-secrecy trade-off is also well captured: A more moderate delay constraint (a smaller $\delta$) yields a larger AEA (a smaller SOP), and vice verser. As expected, the AST scheme remarkably outperforms the NAST one in terms of secrecy, although at the expense of a higher implement complexity.

\begin{figure}[!t]
	\centering
	\includegraphics[width = 3.0in]{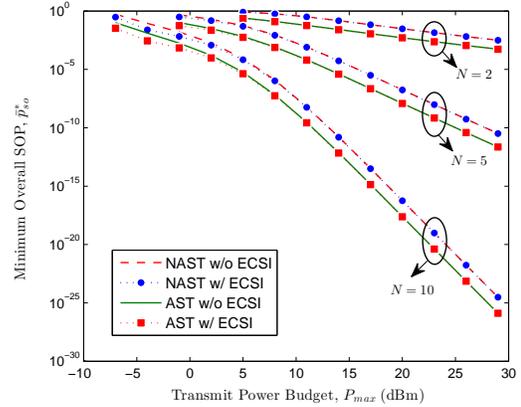}
	\caption{Minimum SOP $\bar p^*_{so}$ vs. $P_{max}$ and $N$, with $\beta_m=1$ and $\delta=0.9$.}
	\label{TSOP}
\end{figure}
Fig. \ref{TSOP} compares the overall SOP under our methods (i.e., maximizing the AEA without ECSI) and under an exhaustive method (i.e.,  minimizing the SOP via \eqref{pso} using the ECSI). We show that the multi-antenna technique plays a significant role in enhancing wireless secrecy. Moreover, the AEA-based SOP approaches the ECSI-based SOP for quite a wide range of $P_{max}$, and they even merge when $P_{max}$ is sufficiently large. This demonstrates the superiority of using the AEA to design secure transmissions when the ECSI cannot be prior known, since the ECSI-irrelevant approach is much easier to implement in practice than those ECSI-based mechanisms.

\section{Conclusions}
This letter considers a secure transmission where secrecy encoding and AN-aided signaling are jointly exploited for anti-eavesdropping when the ECSI is unknown. An ECSI-irrelevant metric AEA is introduced, and both non-adaptive and adaptive schemes are designed to maximize the AEA. Analyses and numerical results show that maximizing the AEA is equivalent to minimizing the SOP, demonstrating the superiority of using AEA to guide secure transmission designs without the ECSI.

\end{document}